\documentclass[journal,10pt]{IEEEtran}

\usepackage[margin=0.66in]{geometry} 
\usepackage{url}
\usepackage{bm}
\usepackage{amsmath}
\usepackage{amssymb}
\usepackage{amsthm}

\usepackage{cite}
\usepackage{algorithm} 
\usepackage{verbatim}
\usepackage{algorithmic} 
\usepackage{bbding}
\usepackage{multirow} 
\usepackage[justification=centering]{caption}
\usepackage{amsmath} 
\usepackage{subfigure}
\usepackage{booktabs}
\usepackage{xcolor}
\usepackage{graphicx} 
\usepackage{epstopdf}
\usepackage{threeparttable}
\usepackage{ragged2e}
\usepackage{amsmath,amsfonts}
\usepackage{algorithmic}
\usepackage{array}
\usepackage[caption=false,font=normalsize,labelfont=sf,textfont=sf]{subfig}
\usepackage{textcomp}
\usepackage{stfloats}
\usepackage{url}
\usepackage{verbatim}
\usepackage{graphicx}
\usepackage {setspace}
\usepackage{pifont}
\usepackage{xurl}

\makeatletter

\newcommand{\Rmnum}[1]{\expandafter\@slowromancap\romannumeral #1@}
\makeatother

\def\BibTeX{{\rm B\kern-.05em{\sc i\kern-.025em b}\kern-.08em
    T\kern-.1667em\lower.7ex\hbox{E}\kern-.125emX}}
\usepackage{balance}

\begin{document}
\begin{spacing}{1}
	
\bibliographystyle{ieeetr}

\title{Space-based Zero-trust Architecture for Securing the Direct-to-Cell Paradigm}

%

\author{Xu~Chen, Jiachen~Sun, Linan~Huang, Wei~Quan, Yi Jing and Xuemin Shen, \IEEEmembership{Fellow,~IEEE}
\thanks{This work was supported by the National Natural Science Foundation of China under Grant 62522127 and 62471274.}
\thanks{X. Chen, J. Sun and Y. Jing are with the Department of Electronic Engineering, Tsinghua University, Beijing 100084, China. (email: chenxu18@tsinghua.org.cn;sjc20@tsinghua.org.cn;jy21@mails.tsinghua.edu.cn).}
\thanks{L. Huang is with the Beijing National Research Center for Information Science and Technology, Tsinghua University, Beijing 100084, China. (email: huanglinan@tsinghua.edu.cn).}
\thanks{W. Quan is with the School of Electronic and Information Engineering, Beijing Jiaotong University, Beijing, China. (email: weiquan@bjtu.edu.cn).}
\thanks{X. Shen is with the Electrical and Computer Engineering Department, University of Waterloo, Waterloo, Canada. (email: sshen@uwaterloo.ca).}

}


\maketitle
\thispagestyle{plain} 
\end{spacing}


%
%

\begin{abstract}
The rapid evolution of Direct-to-Cell (D2C) technology has emerged as a pivotal enabler for ubiquitous 6G global coverage, seamlessly connecting massive numbers of User Equipment (UEs) in remote and underserved areas. However, by bypassing traditional terrestrial satellite access gateways, the D2C paradigm inadvertently renders conventional front-end security perimeters obsolete. Consequently, malicious UEs can directly expose satellites and their core networks to severe security vulnerabilities. To address these critical threats, this article proposes a Space-based Zero-Trust Architecture (SZTA) empowered by on-board satellite computing. Specifically, the proposed SZTA consists of three core components: radio frequency (RF) fingerprint-based authentication, distributed identity management, and behavior-driven dynamic access control. This architecture enables continuous, real-time verification of UE identity authenticity and behavioral legitimacy, effectively mitigating the inherent network-layer security risks of D2C networks. Simulation results demonstrate the robustness, effectiveness, and efficiency of the proposed SZTA. Finally, to facilitate the practical deployment of {the SZTA}, we discuss key technical challenges, propose corresponding solutions, and outline open research issues to guide future investigations.
\end{abstract}

\begin{IEEEkeywords}
Direct-to-Cell, Space-based Zero-trust Architecture, Radio Frequency Fingerprint-based Authentication, Distributed Identity Management, Dynamic Access Control.
\end{IEEEkeywords}

\section{Introduction}


\IEEEPARstart{I}{n} September 2023, China Telecom initiated the world's
first commercial direct-to-cell (D2C) deployment, enabling satellite calls
and messaging on the mass-market Huawei Mate 60 Pro. By March 2026, SpaceX
had deployed 650 D2C satellites, authorized by the U.S. Federal
Communications Commission (FCC) to provide text messaging over T-Mobile's
mobile spectrum \cite{SpaceX2026Rebrands}. The D2C paradigm is thus
rapidly moving satellite mobile communications from an industry-specific
domain toward the mass-consumer market.


D2C offers global coverage, strong anti-jamming capability, and
independence from terrestrial infrastructure, making it particularly
attractive where ground networks are absent. Unlike terrestrial cellular
systems, D2C---also known as satellite direct-to-device (D2D)---employs
satellites as relay nodes to serve mobile user equipments (UEs) directly,
without altering user habits or requiring cross-network protocol
conversion, and is thus expected to underpin the ubiquitous connectivity
envisioned for 6G.


With device miniaturization and low-power design maturing, D2C is expected
to extend beyond smartphones to the broader Internet of Things (IoT)
ecosystem---unmanned systems and IoT devices in offshore, mountainous,
desert, and polar environments can connect directly to satellites
\cite{11429088}. Omdia forecasts 411 million monthly active satellite-D2D
users and \$11.99 billion in revenue by 2030 \cite{Omdia2026Smartphone}.


However, because D2C bypasses satellite access gateways (SAGs), the
traditional SAG-based defense architecture no longer applies: UE traffic
reaches satellites directly and can propagate through inter-satellite
links toward ground control stations, exposing a broader attack surface
and making UE access behavior far harder to monitor. These risks intensify
as the UE population grows.

The primary security issues that the application of D2C may face include the following:

%
%
%

\ding{192} Denial-of-service (DoS) attacks. UE traffic is forwarded
directly to back-end ground control centers for authentication without
prior SAG-based access control, exposing the control servers to DoS
attacks.

\ding{193} Link flooding and routing threats. During authentication,
inter-satellite and satellite--ground links are susceptible to malicious
flooding, and the exposed multi-hop paths invite man-in-the-middle
attacks and route hijacking.

\ding{194} Illegal signal injection. Attackers may inject malicious
instructions or false data through the open satellite channels, jamming
communication or stealing sensitive information, while limited on-board
computing makes real-time detection challenging.

\ding{195} UE identity forgery and privilege abuse. Forged UE identifiers
(e.g., IMEI or MAC addresses) can bypass server-side authentication and
launch large-scale false access requests, exhausting satellite resources
or paralyzing back-end services.

\begin{table*}[!t]
	\centering
	\caption{{Security Risk Evolution, Attacker Classes, and Mitigation Challenges in D2C Networks.}}
	\label{table1}
	\includegraphics[scale=0.54]{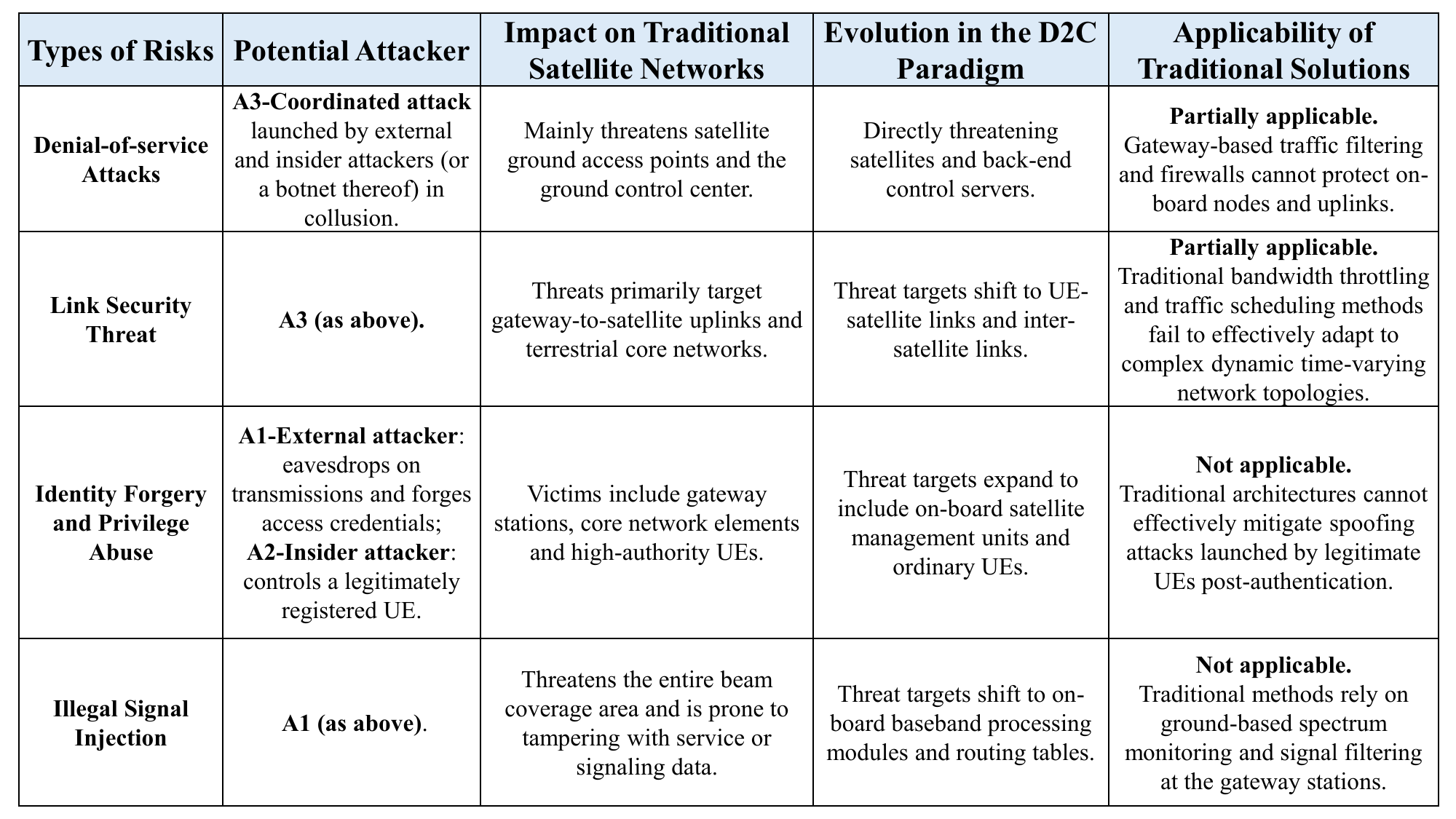}
\end{table*}

{Table~\ref{table1} summarizes the primary security risks associated with the D2C paradigm, the attacker classes that launch them, and the novel challenges that traditional mitigation strategies face in addressing them.}

{Zero-trust technology, featuring dynamic trust evaluation, has proven effective in securing networks with increasingly blurred boundaries, and has recently been extended from enterprise systems to 6G and satellite networks \cite{10288499}. However, existing schemes either anchor security enforcement at terrestrial gateways or the core network \cite{11321204}, allowing malicious traffic to penetrate satellites and inter-satellite links before any defense takes effect; or rely on credential-based verification \cite{farrea2025zero, pokhrel2024poster}, which is ineffective against UEs with forged identifiers or compromised credentials; or lack collaborative authentication across satellites \cite{wei2025bitrust}, such that frequent LEO handovers trigger repeated re-authentication and incur substantial signaling overhead.} To address the aforementioned security challenges, it is imperative to develop a new architecture tailored to the emerging D2C paradigm. The main contributions of this article are summarized as follows.


\begin{itemize}
\item[\large$\bullet$] We propose a space-based zero-trust architecture
(SZTA) for securing D2C communications: security enforcement shifts to
the satellite side, where a lightweight on-board module continuously
verifies UE identity and behavioral legitimacy under non-cooperative
conditions.
\item[\large$\bullet$] We identify the key technical challenges of the
SZTA---non-cooperative identity authentication, distributed identity
management, and dynamic access control---and provide preliminary
solutions.
\item[\large$\bullet$] We verify the security and efficiency of the SZTA
through numerical simulations and discuss open research issues.
\end{itemize}


\section{Space-based zero-trust architecture}

\begin{figure*}
	\centering
	\includegraphics[scale=0.54]{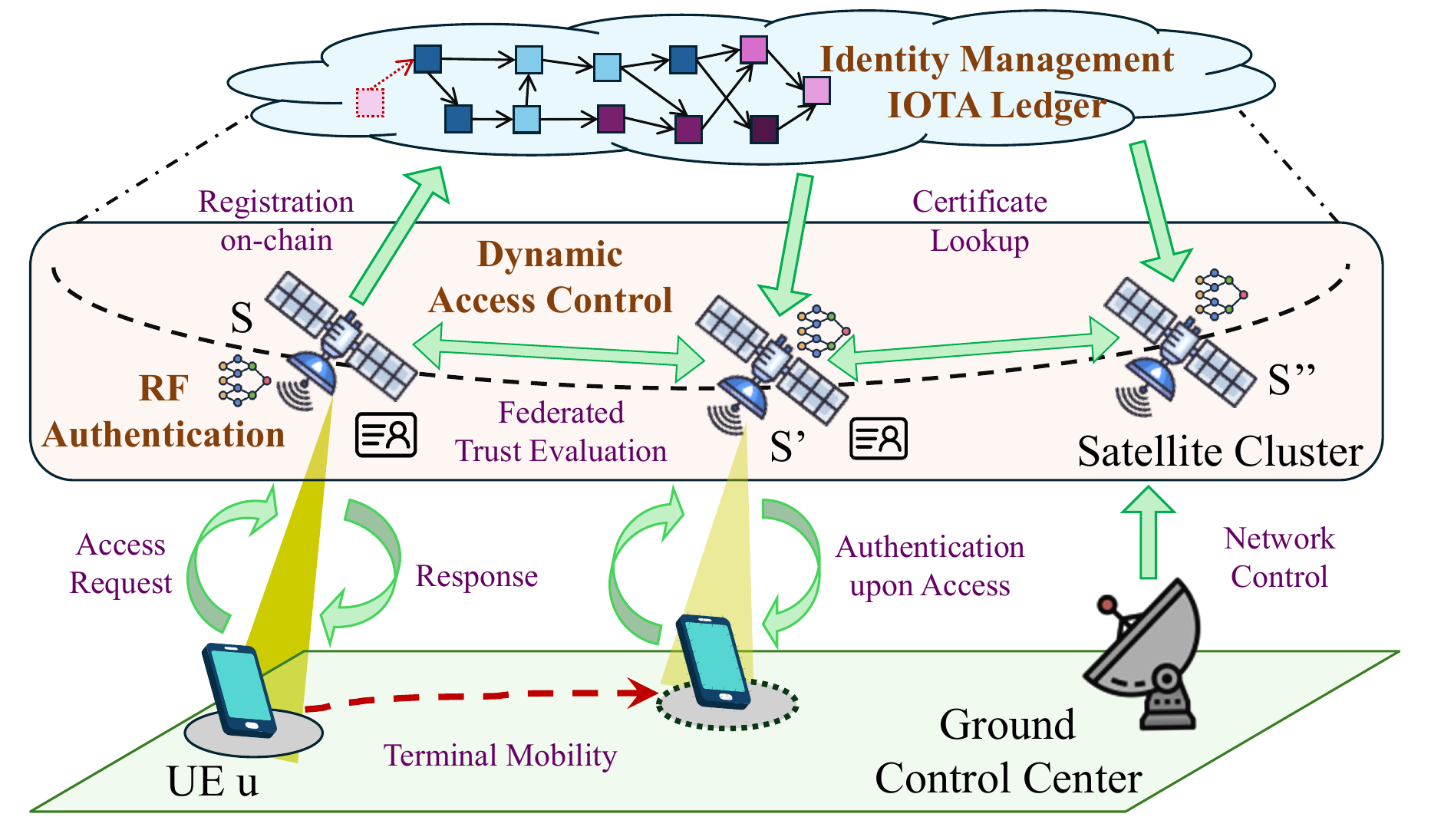}
	\caption{{Space-based zero-trust architecture diagram.}}
	\label{jiagou}
\end{figure*}


The SZTA pursues two objectives: 1) dynamic UE authentication and
least-privilege access control under non-cooperative conditions, without
the single point of failure of centralized back-end authentication; and
2) continuous monitoring of UE behaviors with cross-satellite trust
evaluation that supports dynamic decisions under mobility.

Furthermore, to accommodate the constrained onboard processing architecture, the proposed design adheres to the following principles.

\ding{192} Lightweight design. All security protocols must fit spaceborne
computing and power constraints.

\ding{193} Zero-trust baseline. Every UE is untrusted by default and every
access request is dynamically verified, while minimizing the
certificate-exchange overhead of conventional ZTAs.

\ding{194} Non-cooperative evaluation. Identity and behavior assessment
never relies on information volunteered by the UE, preventing spoofing.

%

To meet these objectives, the SZTA spans the physical to the network layer
and, as illustrated in Fig.~\ref{jiagou}, consists of three modules rely on on-board satellite computing capabilities \cite{10680167}: access authentication verifies UE identity
by combining digital certificates with non-cooperative RF fingerprinting,
establishing the root of trust; identity management maintains UE
identities and behavioral data under high mobility, underpinning trust
evaluation; and dynamic access control makes and enforces access decisions
based on trust results and security policies.

{The operational workflow of the proposed SZTA proceeds as follows, with each stage corresponding to the interactions annotated in
	Fig.~\ref{jiagou}.}
\begin{itemize}
	\item {(\textbf{Initial access and registration})
		When a D2C user $u$ initiates network access, it first sends a request
		to its serving satellite $S$. Satellite $S$ extracts $u$'s RF
		fingerprint, verifies the relevant registration information, generates
		an identity certificate, records the transaction on the IOTA Tangle
		\cite{11313470}, and backs it up at the ground control center.}
	\item {(\textbf{Certificate pre-forwarding})
		After communication is established, $S$ proactively forwards $u$'s
		identity certificate to the next candidate satellite(s) $S'$ based on
		satellite ephemeris and constellation parameters.}
	\item {(\textbf{Authentication upon handover}) Upon
		handover, $S'$ extracts the RF fingerprint of $u$ as an index and
		retrieves the corresponding certificate from its local storage, falling
		back to a ledger lookup when the certificate has not been pre-forwarded
		to $S'$. Once verification succeeds, $u$ is granted access to $S'$ and
		communication continues seamlessly, while $S'$ forwards the certificate
		onward to the next satellite $S''$.}
	\item {(\textbf{Federated trust evaluation})
		When required (e.g., periodically, randomly, or on demand), $S'$
		initiates a trust evaluation request for $u$ to the registration
		satellite $S$, which aggregates the relevant trust factors, computes the
		trust value of $u$, and returns the result.}
	\item {(\textbf{Dynamic access control}) Based on the
		returned trust value and additional security considerations, $S'$ makes
		a comprehensive access decision and enforces dynamic access control
		policies.}
	\item {(\textbf{Revocation}) In the event of
		abnormal conditions---such as account arrears, service termination, or
		detected anomalies---the ground control center issues a revocation
		command to satellite $S$, which revokes the corresponding identity
		certificate and broadcasts the update to other relevant satellites.}
\end{itemize}

\subsection{Radio frequency fingerprint-based authentication}

\begin{figure*}
	\centering
	\includegraphics[scale=0.48]{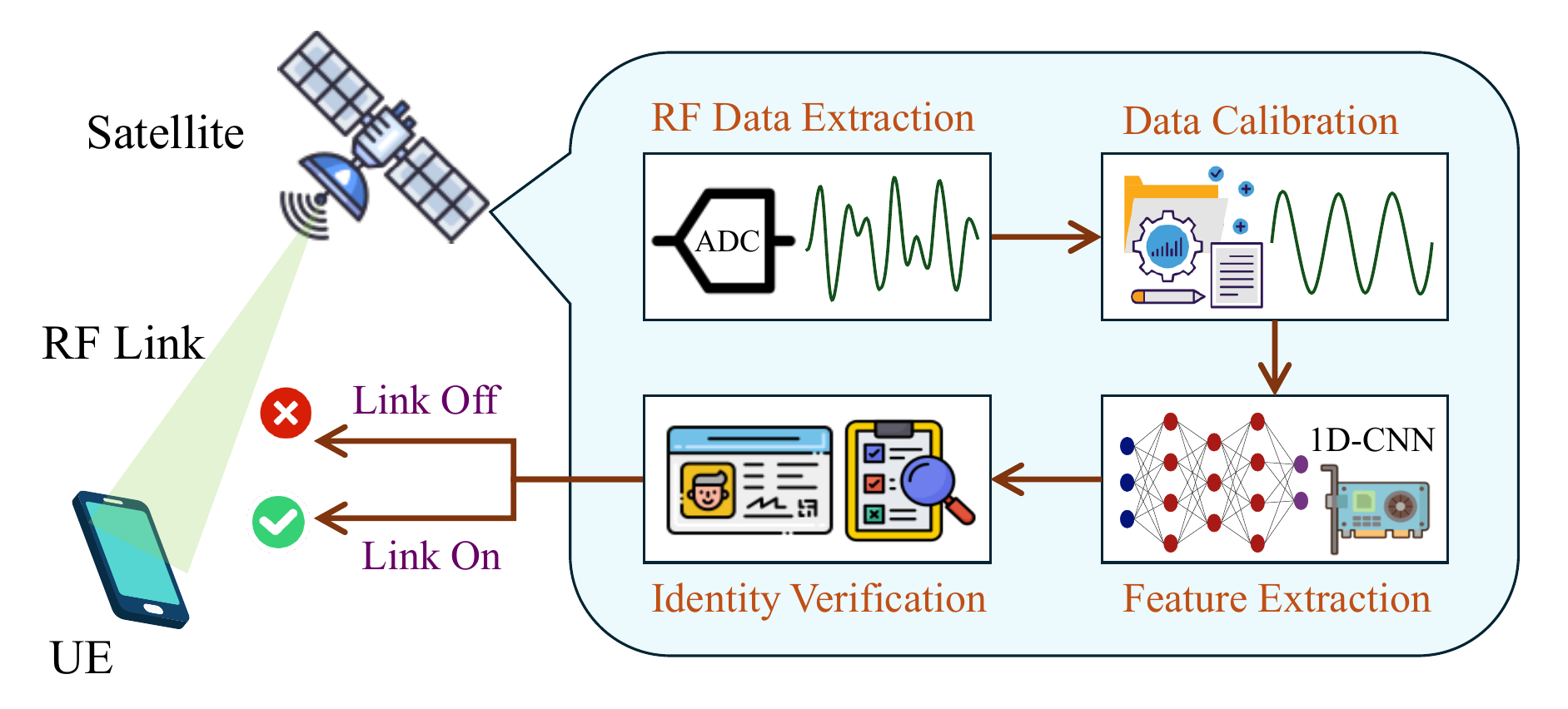}
	\caption{Radio frequency fingerprint feature extraction and UE authentication based on the 1D-CNN.}
	\label{rf}
\end{figure*}


Traditional authentication relies on a centralized Certificate Authority
(CA), which suffers three limitations. First, it is cooperative and
one-shot: once initial authentication completes, it is not re-evaluated
even if the UE later behaves anomalously. Second, the CA must pre-register
credentials for all potential UEs---impractical at 6G scale, and
satellites cannot maintain authentication state for all D2C devices in
advance. Third, CA-based frameworks offer weak defense against identity
spoofing such as forged certificates.


In the SZTA, we instead authenticate UEs via RF fingerprinting: inherent
transmitter hardware impairments, such as nonlinear power-amplifier
distortion and I/Q imbalance, are physically unique and can be captured
from received signals, enabling satellite-side authentication without
extra signaling \cite{ShenTIFS2022}. The challenge is extracting these features under the
rapid UE--satellite relative motion of D2C. We therefore employ a
lightweight one-dimensional convolutional neural network (1D-CNN), whose
temporal convolutions capture local signal patterns, such as peak structures and spectral characteristics. The model
architecture is shown in Fig.~\ref{rf}.


Unlike conventional RFF methods, a feature adaptation module precedes the
1D-CNN: it uses satellite orbital parameters to compensate for
Doppler-induced signal variations, preserving fingerprint fidelity under
high-speed relative motion.


This design enables on-demand authentication in non-cooperative
environments: verification is no longer confined to session initiation but
can be performed continuously, shrinking the attack window, and RF
fingerprints---rooted in intrinsic hardware---are inherently difficult to
forge. Authentication latency is also low: performed on board, it is
dominated by 1D-CNN inference, which is far faster than LSTM-class
alternatives at comparable accuracy \cite{11260469}.

\subsection{Distributed identity management}


Identity underpins any zero-trust architecture. D2C requires a
network-wide identity infrastructure that supports seamless access
anywhere, respects per-satellite resource constraints, and avoids single
points of failure. We therefore build distributed identity management on
RF fingerprints and the IOTA Tangle, a lightweight on-board distributed
ledger.


We adopt a registration--responsibility binding: the satellite that
registers a UE governs its identity thereafter. When UE $u$ first requests
access, serving satellite $S$ combines $u$'s RF fingerprint with its own
identity to generate a unique certificate, which is backed up at the
ground control center but never delivered to the UE---withholding it
prevents leakage and misuse. $S$ subsequently performs all lifecycle
operations on the certificate, including extension, update, and
revocation.

On the satellite side, all registered identity certificates are managed based on the IOTA Tangle architecture. Each satellite maintains a local identity ledger that stores only the certificates of UEs registered by itself. Owing to the relative motion between low Earth orbit (LEO) satellites and ground users, as well as the stochastic nature of UE registration behavior, this identity management mechanism naturally achieves a balanced distribution of identity certificates across satellites, thereby avoiding storage overload at individual nodes. When a UE is connected to the current satellite, its identity certificate is proactively forwarded to the next candidate satellite based on inter-satellite link topology prediction, prior to handover. In exceptional cases where the target satellite fails to receive the certificate in time, it can query the IOTA Tangle using the UE’s RF fingerprint to identify the corresponding registration satellite and request the certificate on-demand. To ensure reliable communication for critical users and high-priority services, the identity states of selected UEs are synchronized across satellites in the constellation via a Gossip protocol. Moreover, the ground control center acts as a coordination node, periodically removing expired data and revoking invalid UE identities.

\begin{figure*}
	\centering
	\includegraphics[scale=0.48]{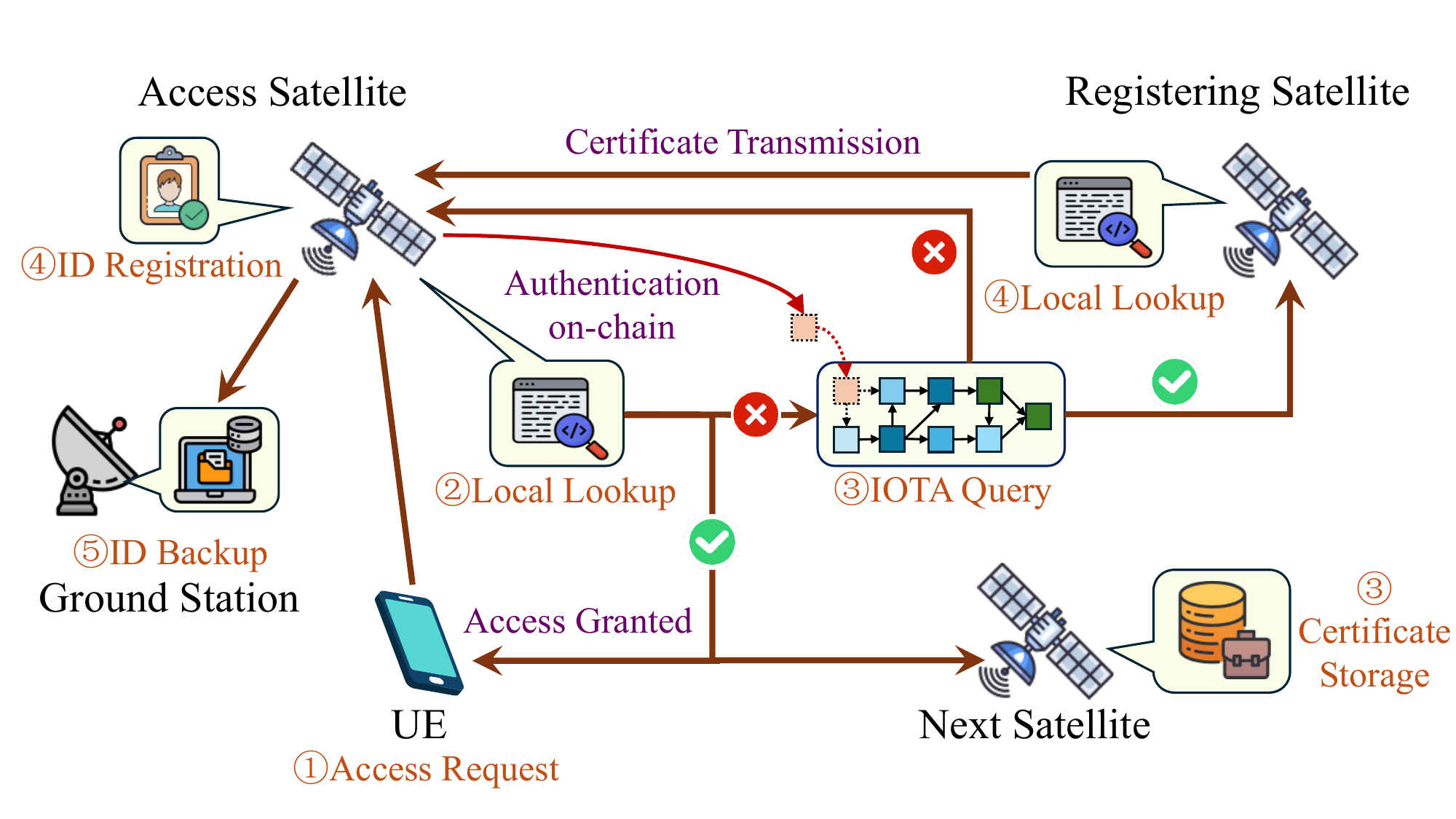}
	\caption{The identity management procedure of a UE in the SZTA.}
	\label{dim}
\end{figure*}

To summarize, the identity management procedure for a UE is illustrated in Fig.~\ref{dim}. When a satellite $S$ receives an access request from a UE $u$, it first extracts the UE's RF fingerprint and searches its local certificate database to determine whether a corresponding identity certificate exists. If a valid certificate is found, authentication and access control are performed accordingly. Otherwise, the satellite queries the IOTA Tangle to identify the UE's registration node and requests the corresponding certificate. If the certificate cannot be retrieved, the UE is treated as an unregistered device. In this case, satellite $S$ generates a new identity certificate based on the extracted RF fingerprint and its own identity information, and subsequently forwards the certificate to the ground control center for secure backup.

\subsection{Behavior-driven dynamic access control}


Admission alone cannot reveal post-access malice; continuous trust
evaluation is therefore central to the SZTA. We build it on federated
learning: by analyzing historical behaviors through inter-satellite
collaboration, the mechanism maintains a dynamic trust score for every UE.


Because a UE's behavioral data are scattered across the satellites it has
visited, modeling requires collaboration: the registration satellite
disseminates an aggregation request via a Gossip protocol, satellites that
previously served the UE upload locally trained behavioral models, and the
registration satellite aggregates them into a composite model against
which current behavior is checked for consistency. Given limited on-board
data and compute, lightweight learners such as random forests are
typically employed. Beyond historical behavioral consistency, the trust score incorporates real-time features---signal stability, spatiotemporal consistency of the signal source, protocol compliance, and energy characteristics---aggregated by a weighted linear combination after normalization.

\begin{figure*}
	\centering
	\includegraphics[scale=0.48]{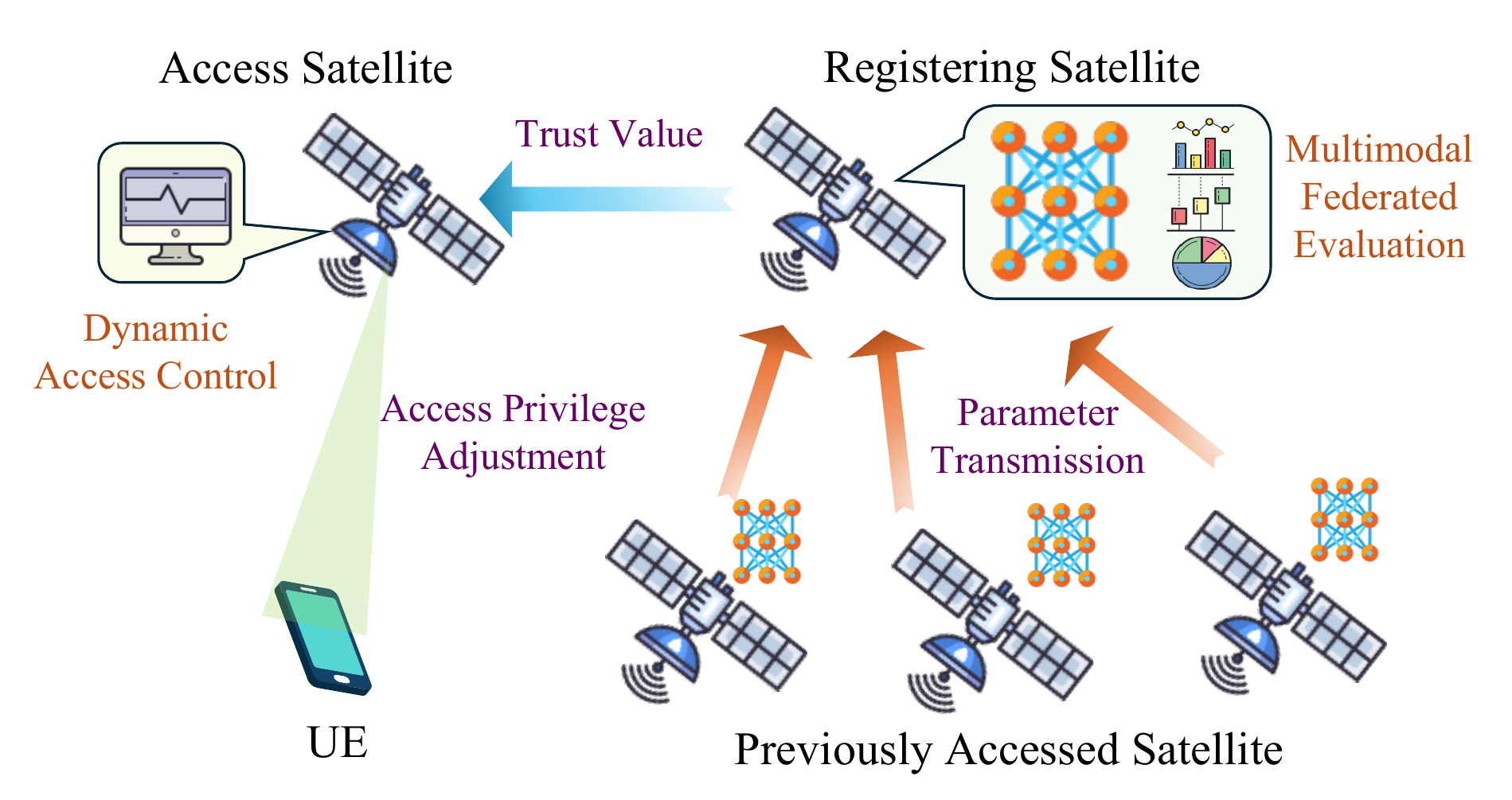}
	\caption{Behavior-driven dynamic access control in the SZTA.}
	\label{bda}
\end{figure*}

Dynamic access control is implemented based on both the evaluated trust score and the current satellite resource status, as {shown} in Fig.~\ref{bda}. Specifically, UE access permissions, such as bandwidth allocation and service priority, are adjusted in real time. In addition, micro-segmentation techniques are employed to isolate high-risk UEs and prevent potential threat propagation. The access control mechanism follows a hierarchical response strategy based on trust levels, enabling differentiated service provisioning and security enforcement.

We now discuss how the proposed SZTA mitigates the aforementioned security threats. For DoS attacks and link security threats, the built-in trust evaluation mechanism of the SZTA can promptly detect bot nodes involved in attacks. It then imposes trust penalties or access restrictions on these bot nodes to mitigate their impact. Furthermore, illegal signal injection attacks can be detected via RF fingerprinting technology. {Because the injected waveform originates from different transmitter hardware, its fingerprint will, with high probability, fail to match the enrolled fingerprint of the legitimate UE.} Regarding identity forgery attacks, the identity authentication mechanism of the SZTA can identify them at both the physical layer and the network layer. Specifically, the physical layer identity is protected by RF fingerprinting technology, while the network layer is secured by the digital certificate authentication mechanism. Therefore, the proposed SZTA effectively defends against network security threats that are inadequately addressed by traditional architectures within the D2C paradigm.

{Furthermore, the proposed SZTA achieves continuous and real-time verification of UE identity authenticity and behavioral legitimacy, thereby mitigating the inherent network-layer security risks of D2C networks through three synergistic dimensions. First, since both RF fingerprint-based authentication and behavior-driven dynamic access control are implemented in a non-cooperative manner—requiring no active coordination from the UE—the SZTA can initiate security assessments at any point in the communication lifecycle, ensuring temporal continuity. Second, the distributed identity management mechanism specifically addresses the challenges of security context correlation and handover under high-mobility conditions, which allows user assessment to be maintained across any location or access satellite, ensuring spatial continuity. Finally, the architecture adopts a lightweight design executed via on-board computing, which bypasses the substantial propagation delays associated with satellite-to-ground links to guarantee the real-time nature of security enforcement. Collectively, the synergy of these three mechanisms enables persistent identity and behavior validation, effectively neutralizing network-layer security threats.}

\section{Performance evaluation}

\begin{figure}[t]
	\centering
	\includegraphics[scale=0.4]{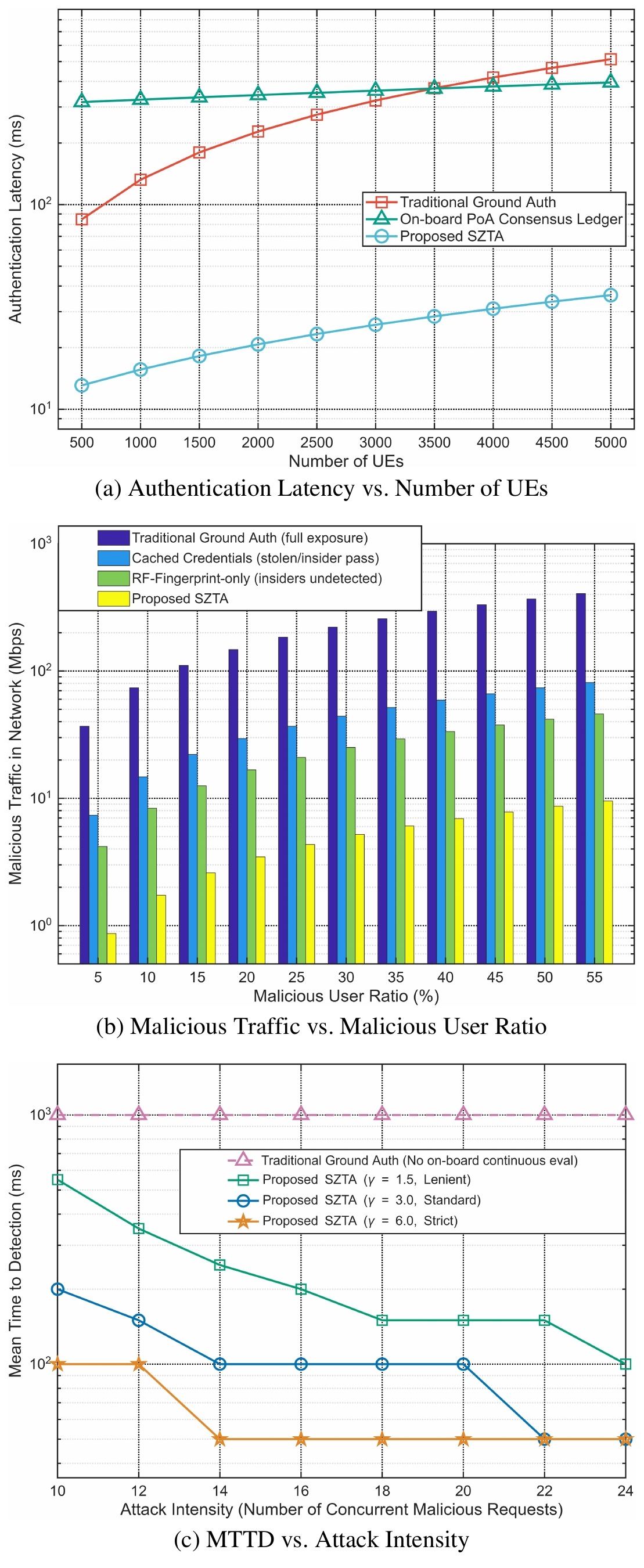}
	\caption{Performance Comparison between SZTA and Traditional Schemes.}
	\label{simulation}
\end{figure}

{We evaluate the SZTA over a Walker-$\delta$ 72/6 constellation at an
altitude of 1000~km (inclination $53^{\circ}$), i.e., 72 satellites evenly
distributed across six orbital planes, served by 6 gateway ground
station. Access requests are routed along
minimum-hop (shortest) paths over the grid ISL topology. The ISL and user up/downlink capacities are 10~Gbps and 500~Mbps, respectively, and each access request occupies 5~KB. On-board authentication is modeled as a 2.4~ms 1D-CNN inference delay for the 114~KB (8-bit quantized) model \cite{giuffrida2021varphi}, plus 1--2~ms of signature-based certificate operations. Each certificate transaction occupies 1.5~KB, with tip selection confined to two ISL hops. The trust score of each UE is initialized to $T_{\max} = 100$ and updated
every 50~ms as
$T \leftarrow 0.9\,T + 0.1\,T_{\max} - \gamma\max(0,\,C - C_{\mathrm{th}})$,
where $C$ is the anomaly measure aggregated from five weighted trust
factors---historical behavioral consistency (0.4), signal stability (0.2),
spatiotemporal consistency (0.2), protocol compliance (0.1), and
energy characteristics (0.1)---$C_{\mathrm{th}} = 5$ is the
tolerance threshold below which no penalty applies, and the penalty
coefficient $\gamma$, swept over $\{1.5, 3.0, 6.0\}$, controls how
aggressively deviations are punished. Scores below $0.5\,T_{\max}$ trigger quarantine with
mandatory re-authentication.}

\subsection{Authentication latency}

{Fig.~\ref{simulation}(a) compares
three schemes: traditional ground authentication, an on-board PoA
consensus ledger, and the proposed SZTA. Ground authentication routes every request to the ground control center. Gateway congestion drives its latency to 513~ms at 5000 UEs. The PoA ledger authenticates on
board, yet before recording a certificate, validator satellites must
exchange multiple rounds of messages to agree on it, costing
0.3~s per write. The SZTA needs no such agreement: in the
Tangle, the serving satellite simply signs the certificate, attaches it
by validating two earlier transactions, and is done---a 1--2~ms local
operation. Consequently, the SZTA verifies every access within
13-36~ms.}

\subsection{Malicious Traffic Bandwidth Consumption}

{Fig.~\ref{simulation}(b) reports the
malicious traffic entering the network as the malicious-UE ratio grows
from 5\% to 55\%, under a mix of 80\% forged devices, 10\% stolen
credentials, and 10\% insiders. Ground authentication exposes the full
405~Mbps at 55\%, since detection waits for the ground-side
audit. Cached credentials admit 81~Mbps: they check only the
credential, so stolen-credential and insider devices pass indefinitely.
RF-Fingerprint-only authentication admits 46~Mbps: it checks only the
hardware, so insiders with genuine devices are never caught. The SZTA
checks both and keeps monitoring afterward, admitting less than 10~Mbps in
total.}

\subsection{Mean Time to Detection (MTTD)}

{The MTTD is compared against varying attack intensities under this framework in Fig.~\ref{simulation}(c). Conventional ground authentication inherently relies on one-time, static authentication, leaving it largely oblivious to latent attacks or privilege abuse occurring post-access. Consequently, the dashed line in Fig.~\ref{simulation}(c) represents a theoretical upper bound, reflecting a prolonged---or even indefinite---detection delay rather than a fixed latency. In contrast, the SZTA perpetually monitors UE behavior. We simulate the response under three distinct penalty policies ($\gamma = 1.5, 3.0$, and $6.0$). As the attack intensity increases, behavioral anomalies rapidly deplete the UE's trust reservoir. Notably, under a stringent policy ($\gamma = 6.0$), an attack with 24 concurrent requests is neutralized within just 50~ms, underscoring the high responsiveness and resilience of the spaceborne zero-trust engine.}

\section{Key technologies and Open issues}


Implementing the SZTA in dynamic D2C environments relies on several
enabling technologies and raises open research issues.

\subsection{Rapid RF fingerprint extraction technology}


Efficient on-board RF fingerprint extraction remains challenging:
satellites must process massive streaming signal data online for fast
classification and identification.

In highly dynamic D2C scenarios, reliable feature extraction further depends on accurate real-time Doppler shift compensation, which requires sub-ppm accuracy. {Dedicated spoofing strategies (replay, transmit-power manipulation, and adversarial waveform crafting) can degrade fingerprint separability, and robust physical-layer authentication under adaptive adversaries remains an open research issue, calling for adversarial training and multi-factor fusion of physical- and network-layer evidence.} Additionally, elastic resource allocation mechanisms are needed to dynamically adapt on-board forwarding resources. To this end, software-defined networking (SDN) provides a promising approach for prioritizing high-trust UE services while ensuring the efficient utilization of limited satellite resources. {Furthermore, a lightweight 1D-CNN inference engine tailored to on-board hardware architectures is required to enable real-time deployment, where the model size typically needs to be constrained to the order of several hundred kilobytes. Since fingerprint characteristics drift with device aging and
constellation evolution, the deployed models must also support secure
over-the-air updates within limited uplink windows, favoring incremental
(differential) model updates and dual-partition images with safe rollback.}

\subsection{Onboard lightweight Tangle protocol}


Implementing a lightweight ledger on resource-constrained satellites is
challenging. An efficient tip selection algorithm should favor
topologically adjacent nodes for validation, minimizing latency and
communication overhead.


Fast fingerprint-indexed certificate retrieval is required for efficient
lookups, and Tangle node functions should be optimized via lightweight
operational modes and hardware acceleration (e.g., FPGAs), keeping
verification energy at a fraction of conventional blockchains.

To further improve scalability, hierarchical consensus mechanisms can be introduced. In this paradigm, the constellation is partitioned into regional clusters, with coordinator satellites responsible for global coordination and subordinate satellites performing local verification, thereby significantly mitigating inter-satellite communication overhead. {Such hierarchical organization also anticipates multi-vendor
constellations: interoperable certificate formats and trust semantics,
aligned with 3GPP NTN standardization, are required so that identities
registered in one operator's cluster remain verifiable across others.}

\subsection{Behavior modeling with space-time constraints}

Accurate behavior modeling in D2C systems requires incorporating spatiotemporal constraints. Satellite ephemeris-based trajectory consistency verification can be employed to validate the legitimacy and plausibility of UE mobility patterns. {In addition, trust factor weights and access policies must remain
remotely updatable after launch, so that operators can recalibrate the detection--false-alarm trade-off as traffic patterns evolve.}

To reduce handover latency, trust evaluation tasks can be proactively delegated to neighboring satellites based on orbit prediction. Federated learning provides a powerful paradigm for distributed behavior modeling, allowing satellite nodes to collaboratively train global trust models without sharing raw data \cite{11744392}. This approach not only mitigates privacy leakage risks but also significantly conserves precious ISL bandwidth resources.

Furthermore, computation offloading can be leveraged to delegate computationally intensive tasks, such as complex behavior analysis, to ground control centers or edge satellites, thereby alleviating the burden on core satellites \cite{11755775}. Energy-aware consensus mechanisms must also be explored to minimize unnecessary computations and prolong the operational lifespan of the constellation.

\section{Conclusion}


In this article, we analyzed the security challenges introduced by the
D2C paradigm and proposed the SZTA, which leverages zero-trust principles
and on-board computing. We detailed its modular composition and
workflow---RF fingerprint-based authentication, distributed identity
management, and behavior-driven dynamic access control---and verified its
system-level effectiveness through constellation-scale simulations. Key
enabling technologies and open issues were outlined; packet-level studies
at mega-constellation scale and transitioning the SZTA into a real-world
solution are left for future work.


\bibliography{document}

\begin{IEEEbiographynophoto}{Xu Chen}
	received his Ph.D. degree in the Department of Electronic Engineering from Tsinghua University in 2022.  His research interests include network management and cyberspace security.
\end{IEEEbiographynophoto}

\begin{IEEEbiographynophoto}{Jiachen Sun}
	received the Ph.D. degree from Tsinghua University, Beijing, China, and is
	currently a postdoctoral researcher with The Hong Kong University of
	Science and Technology, Hong Kong, China, working on
	space-based computing networks.
\end{IEEEbiographynophoto}

\begin{IEEEbiographynophoto}{Linan Huang}
	received the Ph.D. degree from New York University, NY, USA, in 2022, and
	is currently an Assistant Researcher with Tsinghua University, Beijing,
	China, working on satellite network security, game theory, and mechanism
	design.
\end{IEEEbiographynophoto}

\begin{IEEEbiographynophoto}{Wei Quan}
	received the Ph.D. degree from Beijing University of Posts and
	Telecommunications, Beijing, China, in 2014, and is currently a Professor with Beijing Jiaotong University, working on network analytics and AI-driven communications.
\end{IEEEbiographynophoto}

\begin{IEEEbiographynophoto}{Yi Jing} is currently pursuing the
	Ph.D. degree with Tsinghua University, Beijing, China, working on
	satellite communications, resource allocation, and federated learning.
\end{IEEEbiographynophoto}

\begin{IEEEbiographynophoto}{Xuemin (Sherman) Shen} received the Ph.D. degree from Rutgers University, NJ, USA,
	in 1990, and is currently a Professor with the University of
	Waterloo, Canada, working on network resource management.
\end{IEEEbiographynophoto}

\end{document}